\documentclass[%
reprint,
 amsmath,amssymb,
 aps,
 pra,
]{revtex4-2}
\usepackage{graphicx}% Include figure files
\usepackage{amsmath}
\usepackage{amsfonts}
\usepackage{dcolumn}% Align table columns on decimal point
\usepackage{bm}% bold math
\usepackage{braket}
\usepackage[normalem]{ulem}
\usepackage[percent]{overpic}

\begin{document}

\title{Quantum Phase Recognition via Quantum Attention Mechanism}
\author{Jin-Long Chen}% Your name
    \affiliation{Center for Quantum Technology Research and Key Laboratory of Advanced Optoelectronic Quantum Architecture and Measurements (MOE), \\ School of Physics, Beijing Institute of Technology, Beijing 100081, China}
\author{Xin Li}
    \affiliation{Center for Quantum Technology Research and Key Laboratory of Advanced Optoelectronic Quantum Architecture and Measurements (MOE), \\ School of Physics, Beijing Institute of Technology, Beijing 100081, China}
\author{Zhang-Qi Yin}
    \email{zqyin@bit.edu.cn}
    \affiliation{Center for Quantum Technology Research and Key Laboratory of Advanced Optoelectronic Quantum Architecture and Measurements (MOE), \\ School of Physics, Beijing Institute of Technology, Beijing 100081, China}

\date{\today}% It is always \today, today,

\begin{abstract}
 Quantum phase transitions in many-body systems are fundamentally characterized by complex correlation structures, which pose computational challenges for conventional methods in large systems. To address this, we propose a hybrid quantum–classical attention model. 
 This model uses an attention mechanism, realized through swap tests and a parameterized quantum circuit, to extract correlations within quantum states and perform ground-state classification. Benchmarked on the cluster-Ising model with system sizes of 9 and 15 qubits, the model achieves high classification accuracy with less than $100$ training data and demonstrates robustness against variations in the training set. Further analysis reveals that the model successfully captures phase-sensitive features and characteristic physical length scales, offering a scalable and data-efficient approach for quantum phase recognition in complex many-body systems.
\end{abstract}

\maketitle

%\tableofcontents

\section{Introduction}
Quantum phase transitions (QPTs) describe qualitative changes between distinct quantum phases in many-body systems at the ground state, driven by the variation of external parameters, such as coupling strength
\cite{Sachdev1999, vojta2003quantum, sondhi1997continuous}. These transitions are often accompanied by pronounced changes in quantum correlations and entanglement \cite{amico2008entanglement, osborne2002entanglement}, making such correlations natural signatures for phase identification. Understanding QPTs is therefore essential not only for investigating strongly correlated systems and low-dimensional spin models but also for uncovering emergent quantum phases that often lack conventional local order parameters, such as topological or symmetry-protected phases \cite{osborne2002entanglement, osterloh2002scaling}. The complex entanglement structure of quantum many-body systems has stimulated the development of various classical techniques, including mean-field approaches, density functional theories, the density matrix renormalization group (DMRG), and quantum Monte Carlo (QMC) \cite{goldenfeld2018lectures,malpetti2017quantum,kohn1999nobel,schuch2009computational,white1992density,white1993density,becca2017quantum,carlson2015quantum}. While effective in many cases, these methods are limited by correlations, system dimensionality, and computational cost, and they generally fail to detect phases without local order parameters \cite{eisert2010colloquium, troyer2005computational}. This motivates the task of quantum phase recognition (QPR): given a quantum many-body state, can its phase be automatically classified using intrinsic correlations and entanglement patterns?
\begin{figure*}[t]
    \centering
    \includegraphics[width=1.0\linewidth]{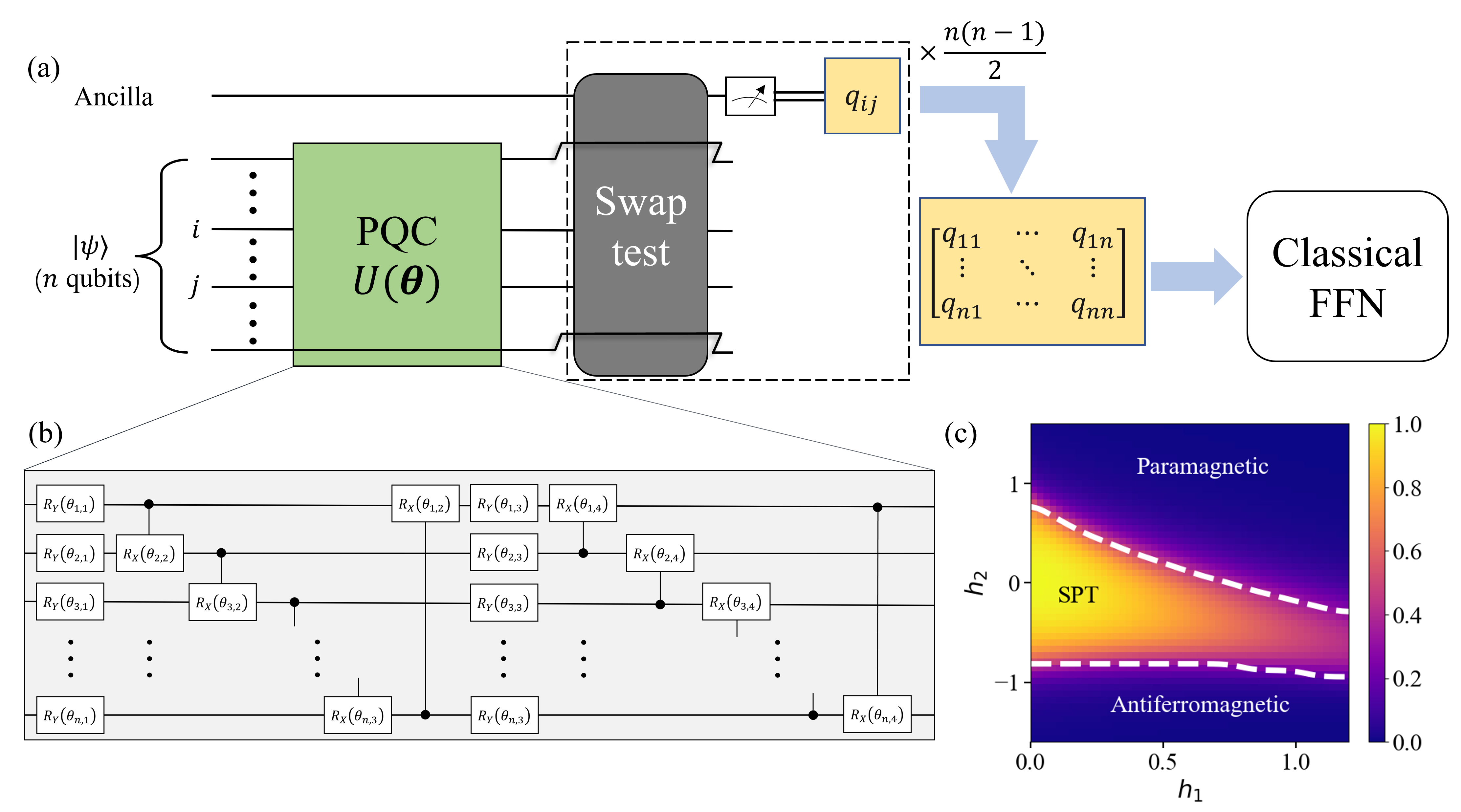}
    \caption{(a) Over framework of attention model. The input state $\ket{\psi}$ is fed into a trainable parametric quantum circuits (PQC) followed by $n(n-1)/2$ swap tests, yielding the attention matrix. The attention values are then fed into a classical feed-forward neural network (FFN) to produce the final output. (b) The single-layer ansatz on $n$ qubits consist of $R_Y$ and $CR_X$. For $l$ layers of circuit on $n$ qubits, the number of parameterised gates (and parameters) is $4nl$. (c) The phase diagram shows the ground-state expectation values of the string order parameter $\langle S \rangle$ in a cluster-Ising model, as a function of the parameters $h_1$ and $h_2$ for a system of size $N=9$. The white dashed lines mark the boundaries between the symmetry-protected topological (SPT) phase and the paramagnetic and antiferromagnetic phases, respectively. }
    \label{fig:model}
\end{figure*}

Quantum computational approaches provide a natural alternative, as they natively operate in the exponentially large Hilbert space \cite{feynman2018simulating, fauseweh2024quantum}. The rapid development of noisy intermediate-scale quantum (NISQ) devices has enabled simulations of interacting quantum systems beyond classical reach. Variational quantum algorithms (VQA), have been proposed to efficiently prepare and probe many-body quantum states \cite{peruzzo2014variational,tilly2022variational}.  Measurement techniques such as classical shadows enable the efficient extraction of physical information from quantum states and have been employed for QPR \cite{huang2020predicting,huang2022provably}. In parallel, classical machine learning (ML) approaches have been applied to many-body QPTs, using sampled configurations, entanglement spectra, or other features of quantum states as inputs, enabling phase identification without explicitly constructing order parameters \cite{carleo2019machine,van2017learning}. These methods can characterize both conventional symmetry-breaking phases and more complex quantum phases, including topological phases \cite{carrasquilla2017machine,schindler2017probing,greplova2020unsupervised,deng2017machine}. However, classical ML relies on classically accessible representations of quantum states, which may become inefficient for highly entangled many-body systems

Quantum machine learning (QML) addresses this limitation by taking encoded quantum states as inputs and learning information through variational or hybrid quantum–classical models \cite{biamonte2017quantum,schuld2015introduction}. Operating directly in Hilbert space allows end-to-end extraction of quantum information \cite{schuld2018supervised,dunjko2018machine}, potentially outperforming classical approaches in expressibility \cite{du2022efficient, abbas2021power, holmes2022connecting}, learning capability \cite{huang2022quantum}, and computational efficiency \cite{liu2021rigorous}. Heuristic QML models, such as quantum neural networks (QNNs) and quantum convolutional neural networks (QCNNs) \cite{farhi2018classification, cong2019quantum}, have been proposed for QPR, leveraging quantum correlations to capture both local and nonlocal features of many-body states. While QNNs often suffer from barren plateaus and high quantum resource demands \cite{mcclean2018barren}, QCNNs trade architectural flexibility for structured locality, limiting their ability to efficiently learn complex or long-range correlations \cite{holmes2022connecting, pesah2021absence, mcclean2018barren}. Recent studies also show that quantum reservoirs, combined with local measurements, can identify quantum phases—including topological transitions—without full state reconstruction \cite{xin2025unsupervised}. 

Inspired by the success of attention mechanisms in classical machine learning for capturing long-range dependencies, recent works have begun exploring analogous mechanisms in quantum architectures \cite{bahdanau2014neural, vaswani2017attention, zhang2025survey}. In quantum many-body systems, phases are typically characterized by intrinsic correlations among constituent particles, which closely parallels the role of attention mechanisms in capturing correlations in classical data. Quantum attention mechanisms therefore hold the potential to integrate information across qubits or quantum states. However, their application to direct quantum data processing remains largely unexplored.

In this work, we introduce a hybrid quantum–classical attention model for QPR. The model leverages the quantum swap test to capture correlations between different qubits, constructing an attention matrix that indirectly encodes global information of the quantum state and enables phase classification. We validate the model via simulations of the Cluster-Ising model on a quantum simulator. Our results show that the model achieves high-accuracy phase classification even with limited training data. Analysis of the learned attention matrices reveals distinctive correlation patterns corresponding to the antiferromagnetic (AFM), symmetry-protected topological (SPT), and paramagnetic phases. Moreover, the model provides interpretable representations of both short-range and long-range correlations, allowing the extraction of effective correlation lengths that serve as order-parameter-like indicators for characterizing phase transitions. These findings not only confirm the effectiveness of the model for QPR but also highlight its potential as a general framework for probing complex correlations in many-body quantum systems.

\section{method}
\subsection{Quautum attention Mechanisms}
Attention mechanisms were originally introduced in classical machine learning to address the challenge of modeling long-range dependencies and global correlations in high-dimensional data \cite{bahdanau2014neural}. Instead of processing all features uniformly, attention assigns adaptive weights to different components of the input, enabling the model to selectively emphasize the most relevant information for a given task. In its canonical form, attention operates by constructing query, key, and value representations and computing their pairwise similarities, which are then normalized to yield an attention matrix that encodes the relative importance of different elements. This weighted aggregation allows attention-based models to efficiently capture nonlocal correlations and has become a central building block in modern architectures such as Transformers \cite{vaswani2017attention}. More broadly, attention can be viewed as a flexible mechanism for learning correlation structures directly from data, making it particularly suitable for tasks where global information and collective behavior play a crucial role.

The attention paradigm has also been extended to quantum computing architectures, most notably in the form of quantum Transformers \cite{zhang2025survey}, which aim to implement or accelerate Transformer-like structures using quantum circuits. Existing approaches to quantum attention can be broadly classified into two technical paradigms. The first employs parameterized quantum circuits to emulate or replace key components of classical attention, such as the generation of queries, keys, and values or the evaluation of attention scores, with the goal of exploiting quantum parallelism and enhanced expressibility on NISQ devices \cite{li2024quantum,xue2024end,zhang2025hqvit,chen2025quantum,he2024training}. The second paradigm is based on quantum linear algebra techniques, which utilize tools such as block encoding, linear combinations of unitaries (LCU) \cite{heredge2025nonunitary,childs2012hamiltonian}, and quantum singular value transformation (QSVT) \cite{gilyen2019quantum} to accelerate matrix operations underlying attention mechanisms \cite{cherrat2024quantum,khatri2024quixer}. While the former is more amenable to near-term hybrid quantum–classical implementations, the latter is primarily of theoretical interest and targets fault-tolerant quantum computers \cite{zhang2025survey}.

Quantum Transformers, which are built on quantum attention mechanisms, are primarily realized within hybrid classical–quantum architectures and have been simulated on various standard benchmark tasks \cite{li2024quantum,zhang2025hqvit,xue2024end,comajoan2024quantum,unlu2024hybrid}. Moreover, several models have been deployed on real quantum hardware and demonstrated performance exceeding classical simulators \cite{he2024training}.

Building on these developments, we now turn to a concrete application of quantum attention mechanisms to QPR. In particular, we introduce a hybrid quantum–classical model in which the attention matrix is constructed directly from measurable qubit–qubit correlations.

\subsection{Hybrid Quantum–Classical Model for QPR} \label{sec:model}
The overall framework of the proposed hybrid quantum-classical model is as shown in Fig.~\ref{fig:model} (a). We first load the quantum states $\ket{\psi}$ into qubits and perform representation learning via a parameterized quantum circuit (PQC) $U(\boldsymbol{\theta})$. To capture correlations among different qubits, a swap-test-based procedure is subsequently applied, in which the effects of exchanging pairs of qubits are quantified and assembled into an attention matrix. This attention matrix is then fed into a classical feed-forward neural network (FFN) to accomplish the classification of multiple quantum phases. The trainable parameters of the PQC, denoted by $\{\boldsymbol{\theta}\}$, together with the weights of the classical FFN, are optimized simultaneously using a gradient-based method in the classical optimizer. The training objective is to minimize the loss function defined between the predicted labels and the true quantum phase labels, following the standard VQA framework. Below, we describe in detail the quantum circuit ansatz, the construction of the swap-test-based attention matrix, and the classical post-processing and training procedure.

We first specify the PQC used for feature mapping. After the preparation of the input state $\ket{\psi}$, the state is passed through a parameterized feature-mapping circuit. In this work, we employ an ansatz $U(\boldsymbol{\theta})$ with strong expressive power and entangling capability~\cite{sim2019expressibility}, whose single-layer $n$-qubit structure is shown in Fig.~\ref{fig:model} (b). Each layer consists of single-qubit $R_Y$ rotation gates and two-qubit controlled-$R_X$ ($CR_X$) gates. The $R_Y$ gates implement local rotations on individual qubits, while the $CR_X$ gates generate entanglement between neighboring qubits. To avoid any bias arising from the distinction between control and target qubits, the roles of control and target are alternated when applying the $CR_X$ gates to adjacent qubit pairs. For a circuit with $l$ layers acting on $n$ qubits, the total number of parameterized gates—and hence trainable parameters—is $4nl$.

Following the feature-mapping stage, correlations between all pairs of qubits are extracted using swap tests (see Appendix~\ref{app:swap test} for details). For the $i$-th and $j$-th qubits, a swap test is performed with the aid of an ancillary qubit, yielding the measurement probability $P_{ij}(0)$ of the ancilla being in the $\ket{0}$ state. From the swap-test measurement outcome, the quantity
\begin{eqnarray}
q_{ij} = 2P_{ij}(0) - 1,
\end{eqnarray}
is obtained, where $q_{ij}$ represents the effect of swapping the $i$-th and $j$-th qubits on the global quantum state, hereby indirectly reflecting the correlation between the two qubits and their relative importance within the many-body system. The resulting attention matrix $\boldsymbol{q}$ is symmetric, since exchanging qubits $i$ and $j$ yields the same outcome, i.e., $q_{ij} = q_{ji}$. For completeness, we define $q_{ii} = 1$. For an $n$-qubit input state, a total of $n(n-1)/2$ swap tests are required to construct the full attention matrix.

The proposed attention mechanism is fundamentally different from the conventional query–key–value (QKV) framework used in classical Transformers. In QKV attention, queries and keys are generated by projecting each input along multiple learned dimensions to compute similarity scores and capture correlations between different inputs \cite{vaswani2017attention}. In our setting, however, we are interested in the intrinsic correlations among subsystems within a single quantum many-body state, and there is no need for multiple projection dimensions. By employing swap tests, we directly construct a symmetric attention matrix that quantifies the effect of exchanging qubits on the global state, providing physically meaningful and interpretable pairwise correlations. Implementing QKV attention would require additional parameter matrices for queries and keys, introducing unnecessary overhead and obscuring the direct connection to the underlying physics. In contrast, our approach avoids redundant embeddings and efficiently captures internal correlations, making it naturally suited for tasks such as QPR.

Once the attention matrix is obtained, it is processed by a FFN for further feature integration and classification. Owing to the symmetry of the attention matrix, only the upper-triangular elements ${\boldsymbol{q}_{ij}}({i<j})$ are used as input to the FFN. Although the FFN can in principle be implemented using either quantum or classical resources, we adopt a classical implementation to reduce quantum resource requirements. In this work, the FFN consists of a single fully connected layer followed by a nonlinear activation function $\sigma(\cdot)$, yielding
\begin{eqnarray}
y = \sigma(\boldsymbol{w}^T \boldsymbol{q} + b),
\end{eqnarray}
where $\boldsymbol{w}$ and $b$ denote the trainable weights and bias, respectively. The activation function $\sigma$ is chosen as the sigmoid function for binary classification or the softmax function for multi-class classification. The output $y$ represents the predicted phase label for the input quantum state $\ket{\psi}$. During training, we employ the cross-entropy loss function, and both the quantum circuit parameters and the classical neural network parameters are optimized concurrently by a classical optimizer.

After introducing the architecture of the hybrid quantum-classical attention model, we now analyze its computational complexity with respect to the system size and circuit depth. For an $n$-qubit system with $l$ circuit layers, the feature-mapping circuit contains $\mathcal{O}(nl)$ parameterized gates, leading to linear scaling in both the system size and the circuit depth. The construction of the attention matrix requires swap tests between all pairs of qubits, resulting in a quadratic complexity $\mathcal{O}(n^2)$. The subsequent classical FFN processes the upper-triangular elements of the symmetric attention matrix with a computational cost of $\mathcal{O}(n^2)$. Overall, the total complexity scales as $\mathcal{O}(n^2 + nl)$, which is dominated by the quadratic term for large $n$. In particular, in this work we adopt a shallow circuit with $l=1$, where $l \ll n$, such that the overall complexity effectively reduces to $\mathcal{O}(n^2)$.

\section{Results and Discussion}\label{sec:results}
As a representative quantum many-body system hosting both SPT order and conventional symmetry-breaking phases, we choose the cluster–Ising model as a benchmark, where the SPT phase can not be characterized by any local order parameter. The Hamiltonian is given by \cite{smacchia2011statistical}: 
\small
\begin{eqnarray}
H = -J \sum_{i=1}^{N-2} Z_i X_{i+1} Z_{i+2} 
    - h_1 \sum_{i=1}^{N} X_i 
    - h_2 \sum_{i=1}^{N-1} X_i X_{i+1},
\label{eq:hamiltonian}
\end{eqnarray}
where $X_i$ and $Z_i$ denote the Pauli operators acting on the spin at site $i$, and $h_1$, $h_2$, and $J$ are the model parameters. The ground states of this Hamiltonian can belong to a SPT phase, a paramagnetic phase, or an AFM phase, depending on the values of $\{h_1/J, h_2/J\}$. For simplicity, we set the coupling constant $J = 1$ throughout this study. In the special case $h_2 = 0$, the Hamiltonian becomes exactly solvable via the Jordan-Wigner transformation~\cite{Sachdev1999}.
\begin{figure}
    \centering
    \includegraphics[width=\linewidth]{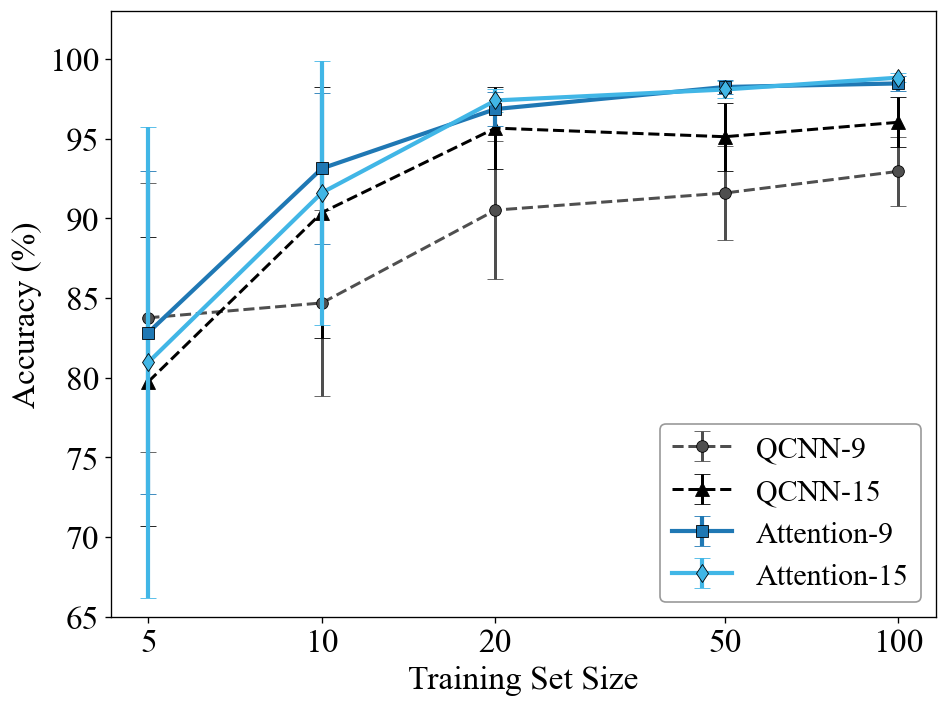}
    \caption{Classification accuracy of QCNN and attention-based models with 9 and 15 qubits as a function of the training set size. Each data point is obtained by averaging over 10 independent runs with randomly selected training datasets. Error bars indicate the 95\% confidence interval.}
    \label{fig:Accuracy}
\end{figure}

A key feature of this Hamiltonian is its $\mathbb{Z}_2 \times \mathbb{Z}_2$ symmetry, which arises from its commutation with both the even ($\prod_i X_{2i}$) and odd ($\prod_i X_{2i+1}$) parity operators. This symmetry class is the same as that of the $S=1$ Haldane phase~\cite{verresen2017one,haldane1983nonlinear}. The three phases, and in particular the SPT phase, can be distinguished by the expectation value of the corresponding string order parameter~\cite{pollmann2012detection}, defined for an odd number of spins $N$ as
\begin{equation}
    S = Z_1X_2X_4\cdots X_{N-3}X_{N-1}Z_N,
\end{equation} 
\begin{figure*}
    \centering
    \includegraphics[width=\linewidth]{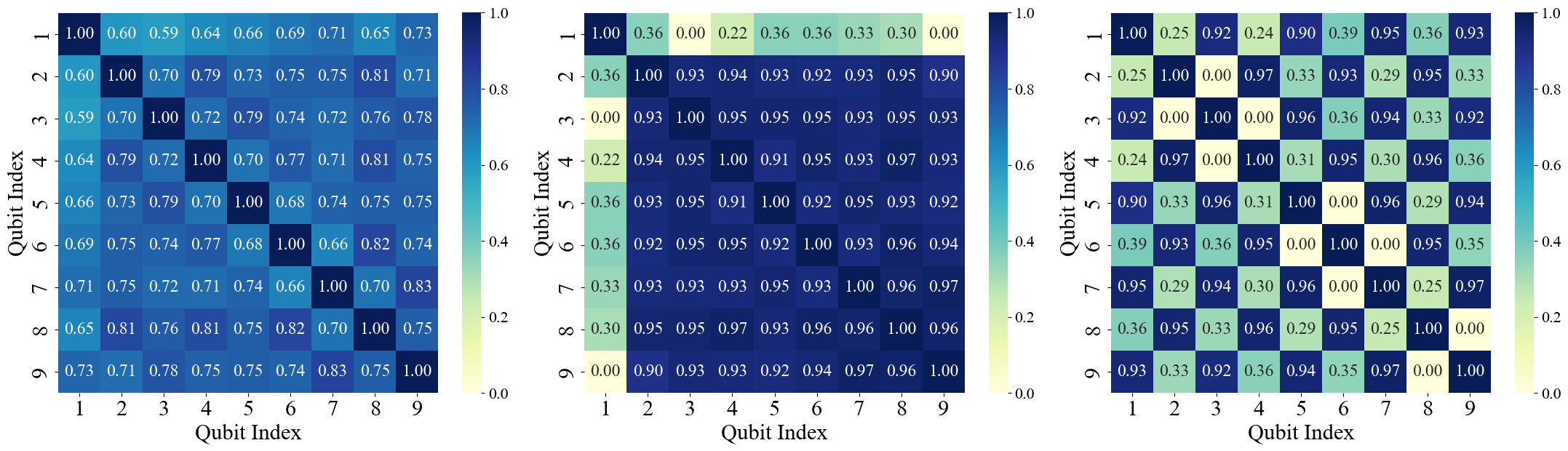}
    \caption{Heatmaps of attention matrices revealing distinct correlation patterns in different quantum phases. The axes correspond to qubit indices, and the color intensity represents the correlation strength. SPT phase: uniformly large matrix elements, indicating nonlocal entanglement. Paramagnetic phase: partially decoupled pattern, with one qubit effectively decoupled from the rest, reflecting dominant local polarization. Antiferromagnetic phase: staggered structure with alternating strong and weak correlations, characteristic of Néel-type order.}
    \label{fig:matrix}
\end{figure*} 

\begin{figure}
    \centering
    \includegraphics[width=1\linewidth]{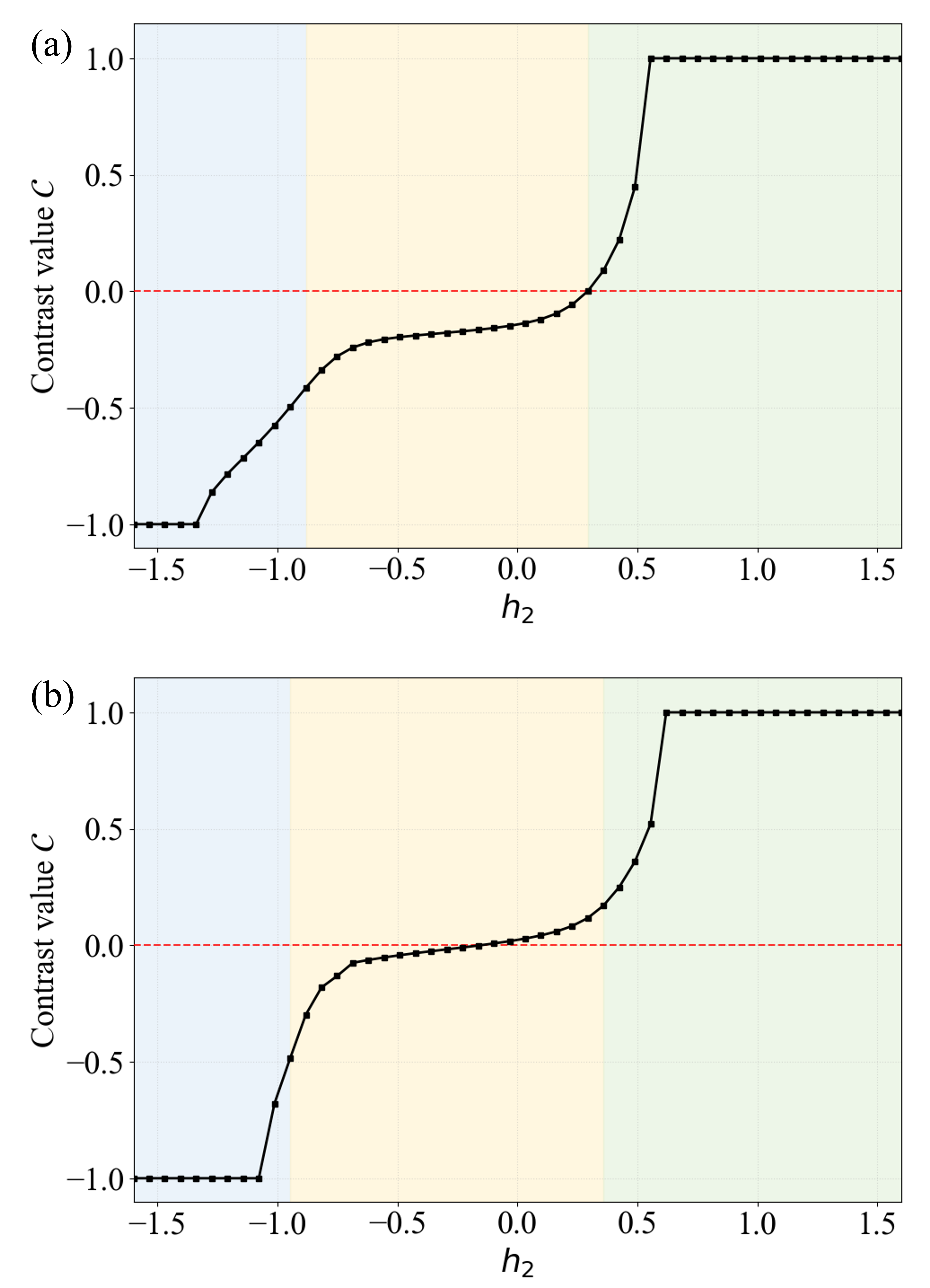}
    \caption{Contrast value $ C = (q_\text{near}-q_\text{long})/(q_\text{near}+q_\text{long})$ as a function of the control parameter $h_2$ at fixed $h_1 \approx 0.39$ for (a) 9-qubit and (b) 15-qubit systems. From left to right, the system exhibits the AFM, SPT, and paramagnetic phases. Sharp sign changes of $C$ clearly delineate the phase boundaries, providing an order-parameter-like signature of the quantum phase transitions.}
    \label{fig:Contrast value}
\end{figure}

We consider quantum spin systems of different sizes, specifically $N=9$ and $N=15$ qubits. For each system size, a dataset of $2500$ ground states is generated by exactly diagonalizing the Hamiltonian in Eq.~\eqref{eq:hamiltonian}, with Qiskit \cite{javadi2024quantum}. The phase label of each ground state is determined from the expectation value of the string order parameter $\langle S \rangle$. Based on this quantity, we construct the phase diagram for the $N=9$ system, as shown in Fig.~\ref{fig:model} (c). Due to finite-size effects, the phase transitions appear as smooth crossovers rather than sharp boundaries. The crossover lines (white dashed lines) are identified by locating the maxima of the second derivative of the energy expectation value $\langle H \rangle$ with respect to $h_2$~\cite{Sachdev1999}.

Building on the generated datasets, each data point is represented as a pair $(\ket{\psi(h_1,h_2)}, b)$, where $\ket{\psi(h_1,h_2)}$ denotes the ground state of the parameterized cluster-Ising Hamiltonian $H(h_1,h_2)$, and $b$ corresponds to its phase label. The training set is then constructed by randomly sampling such data pairs from the full dataset.

Given an unknown ground state $\ket{\psi(h_1,h_2)}$, the task is to determine its corresponding quantum phase. This classification is performed using the quantum machine learning model introduced in Sec.~\ref{sec:model}, which is trained to extract phase-relevant features and assign the appropriate phase label.

We first evaluate the model's ability to reconstruct the phase diagram using a limited number of training samples under ideal conditions with a quantum simulator, implemented using PennyLane \cite{bergholm2018pennylane}. Figure~\ref{fig:Accuracy} presents the classification accuracy over the entire phase space, averaged over ten independent runs for each training set size, with error bars indicating the 95\% confidence interval. The accuracy improves rapidly as the training set size increases and saturates at approximately 20 samples. Remarkably, the 9-qubit system achieves up to 98\% accuracy with only 20 training pairs, while the 15-qubit system exhibits similar behavior. These results demonstrate the model's data efficiency and strong generalization capability, which is particularly advantageous for larger systems where generating labeled ground states is computationally expensive \cite{caro2022generalization}.

The model also demonstrates robustness against statistical fluctuations and variations in the training set. Relatively small error bars indicate consistent performance across independent runs. In the low-data regime, particularly when the training set exceeds 50 samples, the standard deviation remains below 1\%, demonstrating that the learned decision boundaries are stable and largely insensitive to the specific choice of training samples. Compared with QCNN, the attention-based model achieves higher recognition accuracy and improved stability under similar conditions.

\subsection{Phase-dependent attention patterns and contrast measure}
To investigate the phase-specific information captured by the model, attention matrices are extracted from the trained network for the ground states in each phase, with a representative example shown in Fig.~\ref{fig:matrix} (see Appendix~\ref{app:Attention Matrices} for  more examples). In the SPT phase, the matrix elements are nearly uniform, reflecting nonlocal entanglement characteristic of topological order. In contrast, the paramagnetic phase displays a partially decoupled pattern, with one or a few qubits weakly correlated with a the rest, indicating dominant local polarization. For the antiferromagnetic phase, the matrix displays a staggered structure with alternating strong and weak correlations, consistent with Néel-type ordering. These distinct structures clearly differentiate the phases and demonstrate the model’s ability to capture intrinsic features of each quantum state.

The phase-dependent patterns in the attention matrices are clearly apparent. To make these differences quantitative and comparable across parameters, we extract a simple metric from the attention matrices and track their evolution across the phase diagram. We define contrast value as
\begin{eqnarray}
    C = \frac{q_\text{near}-q_\text{long}}{q_\text{near}+q_\text{long}}
\end{eqnarray}
where $q_\text{near}$ and $q_\text{long}$ are the elements of the attention matrices, representing the correlations between the qubit with significantly reduced correlations and its nearest-neighbor qubit, and between the same qubit and the distant qubit, corresponding to nearest-neighbor and long-range correlations, respectively. For example, consider the 9-qubit system shown in Fig.~\ref{fig:matrix}, where the nearest-neighbor correlation is measured between qubits 1 and 2, and the long-range correlation is captured between qubits 1 and 9. In this case, $q_\text{near} = q_{12}$ and $q_\text{long} = q_{19}$.  With $h_1 \approx 0.39$ fixed, we study its evolution as a function of the parameter $h_2$ with the corresponding results presented in Fig.~\ref{fig:Contrast value}. It can be seen that in the antiferromagnetic phase, long-range correlations dominate; in the paramagnetic phase, nearest-neighbor interactions are more pronounced; while in the SPT phase, long-range correlations are slightly stronger than nearest-neighbor ones, though the two are roughly comparable. Sharp sign changes in 
$C$ clearly indicate phase boundaries, providing an order-parameter-like signal for the transitions. 
\begin{figure}
    \centering
    \includegraphics[width=\linewidth]{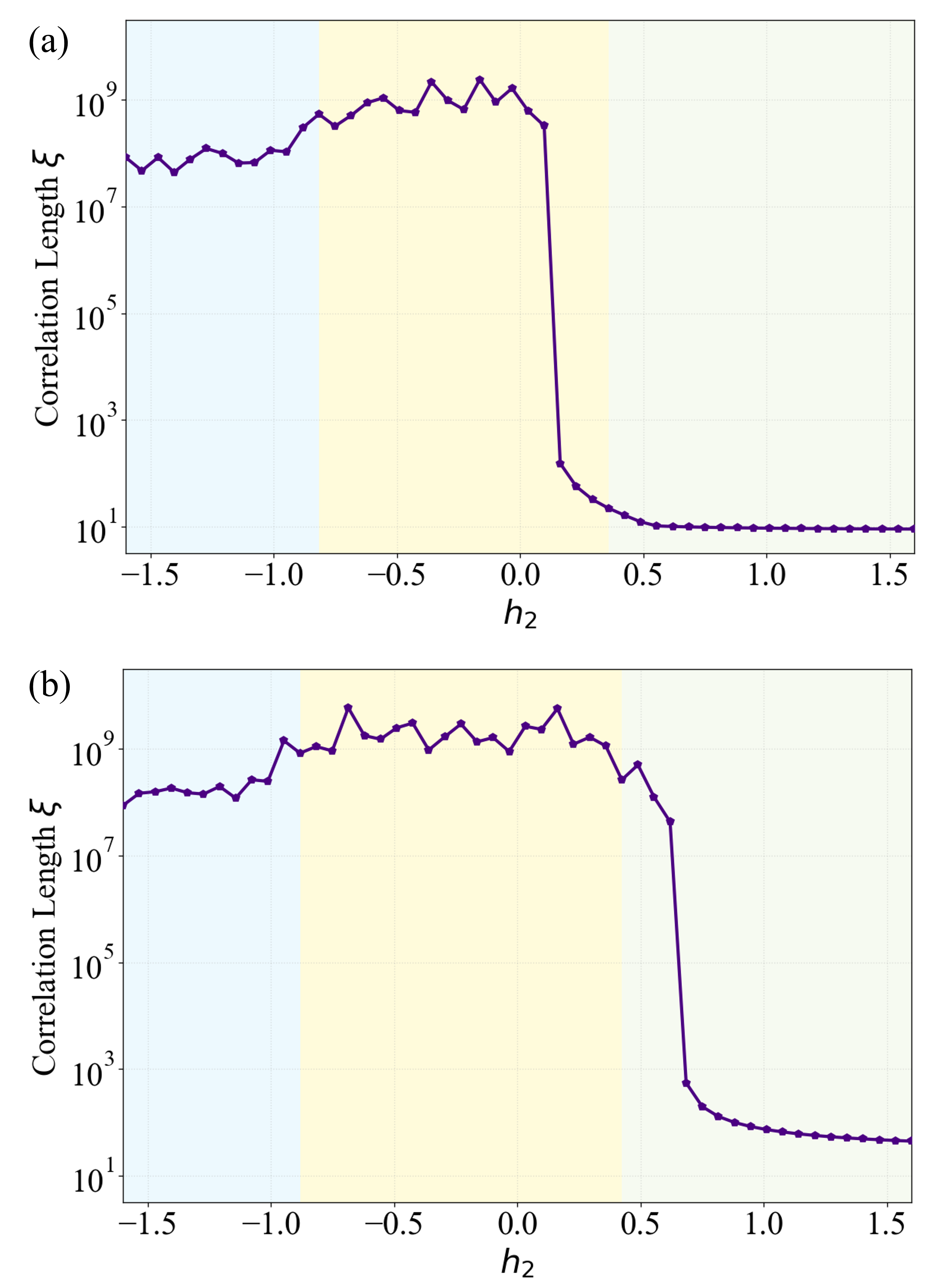}
    \caption{Effective correlation length $\xi$ from the attention matrix versus $h_2$ at fixed $h_1 \approx 0.39$ for (a) 9-qubit and (b) 15-qubit systems. From left to right, the system exhibits the AFM, SPT, and paramagnetic phases. In the AFM phase, $\xi$ is one order smaller than that in the SPT phase, indicating that while long-range correlations exist, they are less dominant than the non-local topological order in the SPT phase; in the SPT phase, $\xi$ is strongly enhanced and reaches its peak, reflecting nonlocal topological order; in the paramagnetic phase, $\xi$ is in the order of $10$, consistent with localized correlations.}
    \label{fig:correlation length}
\end{figure}

\subsection{Distance-dependent correlations and effective length scales}
While the contrast value $C$ provides a concise measure of the competition between nearest-neighbor and long-range correlations, it probes only two representative length scales in the system and therefore cannot capture the continuous spatial evolution of correlations. To obtain a more complete picture of the spatial correlation structure learned by the model, we introduce a correlation strength function extracted from the attention matrix. 

In practice, the construction of correlation strength function is motivated by the observation that, in certain phases such as the paramagnetic phase, some qubits exhibit significantly weaker correlations with the rest of the system. We therefore select the first (or last) such weakly correlated qubit as a reference and restrict the attention matrix to a submatrix consisting of elements located on one side of this reference qubit (either after or before it). Within this restricted sector of the attention matrix, we define the distance-dependent correlation strength 
\begin{eqnarray}
    f(r) = \frac{1}{N-r}\sum_{i=1}^{N-r}q_{i,i+r}
\end{eqnarray}
where $N$ denotes the dimension of the selected attention submatrix, and $r$ is the distance between the two qubits within this submatrix.

Notably, our correlation strength function differs from conventional two-point correlators used in quantum many-body physics \cite{malpetti2016quantum, liu2023model}. While the latter quantify physical fluctuations of specific observables, the correlation strength function reflects the learned importance assigned by the neural network to different spatial separations. It effectively serves as a proxy for the system’s entanglement structure, without prior knowledge of the optimal order parameter. The quantity $f(r)$ can be interpreted as an effective spatial correlation strength inferred by the model. Large values of $f(r)$ indicate that the model assigns significant importance to correlations between sites separated by a distance $r$, whereas a rapid decay of $f(r)$ reflects predominantly local correlations.

To further quantify the spatial decay of correlations learned by the model, we fit $f(r)$ to an exponential form, $f(r) \sim e^{-r/\xi}$ with $\xi$ interpreted as an effective correlation length. The extracted correlation length $\xi$ as a function of the control parameter $h_2$, with $h_1 \approx 0.39$ fixed for both 9- and 15-qubit systems, is shown in Fig.~\ref{fig:correlation length}. Distinct behaviors are observed across different quantum phases. In the AFM phase, $\xi$ is on the order of $10^8$, but it is much smaller compared to the SPT phase, indicating that the model predominantly captures short- to intermediate-range correlations. In the SPT phase, the effective correlation length $\xi$ is significantly enhanced to around $10^9$, reflecting nonlocal correlations characteristic of topological order. By comparison, in the paramagnetic phase, $\xi$ rapidly decreases toward $10$, consistent with the dominance of local correlations and the absence of long-range order. Notably, the largest values of $\xi$ in the SPT phase are signatures of negligible decaying correlations rather than true divergences. The resulting $\xi$ shows a clear dependence on the quantum phase, indicating that the attention mechanism captures not only phase information but also characteristic physical length scales.

\section{CONCLUSION}
In this work, we have introduced a hybrid quantum–classical attention model within the VQA framework, employing the quantum swap test to enable QPR in many-body systems. Motivated by the central role of quantum correlations and entanglement in characterizing QPTs, the proposed model learns internal correlation structures of ground states with nontrivial phases. Validation on a quantum simulator demonstrates that the model achieves high classification accuracy even with a limited number of labeled samples, indicating strong generalization capability in the low-data regime \cite{caro2022generalization}. Moreover, analysis of the attention matrices learned by the model reveals clear distinct differences among various quantum phases in terms of nearest-neighbor correlations, long-range correlations, and correlation lengths. These findings uncover the critical information captured by the model and provide a reliable basis for predicting phase labels. 

In classical Transformers, visualizing attention matrices often provides meaningful insights into the structure of the input data \cite{caron2021emerging}. Following this idea, the quantum attention matrices learned by our model can also be directly visualized, revealing the system’s internal correlation patterns. These visualizations capture key physical properties, including nearest-neighbor correlations, long-range correlations, and effective correlation lengths, enabling clear distinction among AFM, SPT, and paramagnetic phases. This approach not only demonstrates the model’s ability to extract meaningful quantum phase features but also provides an intuitive and practical way to interpret model decisions and understand the underlying correlation structures in complex many-body systems.

Importantly, the proposed framework is not restricted to the specific cluster-Ising model studied here. Since it relies on intrinsic correlations encoded in quantum states rather than explicit order parameters, it is readily extendable to a broad class of quantum many-body systems, including models exhibiting topological or symmetry-protected phases where conventional diagnostics are absent. In this sense, quantum attention provides a versatile and physically motivated approach for probing complex many-body phenomena through internal correlation structures.

Nevertheless, the present study is limited to small system sizes and idealized quantum simulators. Scaling the approach to larger systems and implementing it on near-term quantum hardware will require careful consideration of quantum resource costs, noise, and decoherence, which may affect both classification accuracy and the extraction of correlation features. Future work may focus on resource-efficient circuit designs, noise-resilient training strategies, and experimental demonstrations on quantum devices \cite{shao2024simulating, preskill2018quantum}. Extending the framework to higher-dimensional systems and to models lacking well-defined local order parameters may further establish quantum attention as a powerful and versatile tool for investigating quantum phase transitions in complex quantum matter.

\begin{acknowledgments}
We thank the developers of Qiskit and PennyLane, which we used for ground-state searches and quantum circuit simulations, and we also acknowledge OpenAI’s GPT for assistance with manuscript writing and polishing. This work was supported by the Beijing Institute of Technology Research Fund Program under Grant No.2024CX01015.
\end{acknowledgments}

\appendix
\section{Swap Test}\label{app:swap test}
\begin{figure}[h]
    \centering
    \includegraphics[width=0.5\linewidth]{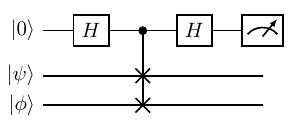}
    \caption{The swap test circuit}
    \label{fig:swap test}
\end{figure}

The swap test is a widely employed quantum operation for determining the similarity between two quantum states. Moreover, it serves as an effective tool for characterizing the change resulting in the global system under the exchange of two states or qubits. A swap test is implemented through a simple interferometric circuit consisting of an ancilla qubit and two target registers holding the states to be compared, which is shown in Fig.~\ref{fig:swap test}.
\begin{figure*}
    \centering
    \includegraphics[width=\linewidth]{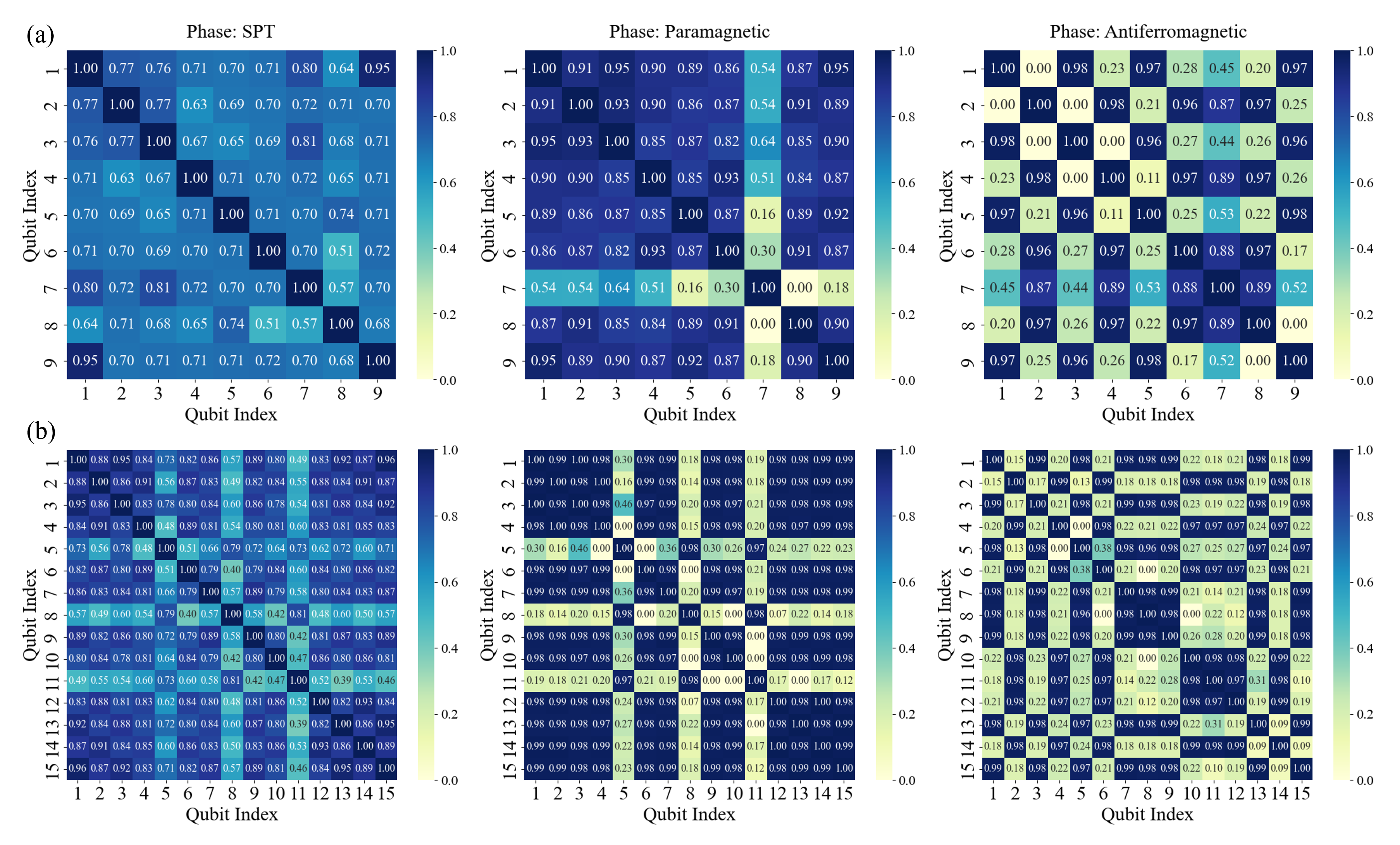}
    \caption{Measured attention matrix for the qubit register for  9-qubit and (b) 15-qubit system. Each element represents the pairwise correlation coefficient between the states of two qubits (indexed by rows and columns). Coefficients near 1 indicate strong correlation, while values near zero suggest independence.}
    \label{fig:additional matrix}
\end{figure*}

We regard the target qubits as the entire system, i.e., $\ket{\Phi}\equiv\ket{\psi}\ket{\phi}$. The ancilla is first initialized in $\ket{0}$ and subjected to a Hadamard gate, creating a coherent superposition that controls the subsequent operation. 
\begin{eqnarray}
    \ket{\Psi_1}=\frac{1}{\sqrt{2}}(\ket{0}\ket{\Phi}+\ket{1}\ket{\Phi}),
\end{eqnarray}

A controlled-SWAP gate is then applied, where the ancilla qubit governs the exchange of the two target states. 
\begin{eqnarray}
    \ket{\Psi_2}=\frac{1}{\sqrt{2}}(\ket{0}\ket{\Phi}+\ket{1}\ket{\Phi_{sw}}),
\end{eqnarray}
where $\ket{\Phi_{sw}}$ is the state on the target register after performing the swap operation.Another Hadamard gate is performed on the ancilla before measurement. The final state is 
\begin{eqnarray}
    \ket{\Psi_3}=\frac{1}{2}(\ket{0}\ket{\Phi}+\ket{\Phi_{sw}})+\ket{1}(\ket{\Phi}-\ket{\Phi_{sw}}),
\end{eqnarray}
Then the probability that the measurement yields 0 for the auxiliary qubit is 
\begin{align}
  \mathrm{Prob}(0) &= |\frac{1}{2}(\ket{\Phi}+\ket{\Phi_{sw}})|^{2}\notag\\
    &=\frac{1}{2}(1+\langle \Phi | \Phi_{sw} \rangle).
    \label{eq:prob}
\end{align}

From Eq. \eqref{eq:prob}, we obtain
\begin{eqnarray}
\langle \Phi | \Phi_{sw} \rangle = 2\mathrm{Prob}(0)-1.
\label{eq:q}
\end{eqnarray}

Thus, the change induced by the swap operation can be quantified by measuring the probability of the auxiliary qubit.

\section{Additional Attention Matrices}\label{app:Attention Matrices}
As discussed in the Sec. \ref{sec:results}, the attention matrices obtained by the hybrid quantum-classical attention model exhibit distinct correlation patterns across different quantum phases. In the SPT phase, the matrix elements are uniformly large, indicating the presence of nonlocal entanglement. The paramagnetic phase is characterized by a highly asymmetric pattern, where one or a few qubits are effectively decoupled from the rest, reflecting dominant local polarization. In contrast, the antiferromagnetic phase displays a staggered structure with alternating strong and weak correlations, which is characteristic of Néel-type order.

In this Appendix, we present additional attention matrices obtained from independently trained networks for various ground states across the phase diagram, as shown in Fig.~\ref{fig:additional matrix}. These additional matrices complement the results shown in the main text, illustrating a degree of reproducibility of the learned correlation structures across independent training runs. Each matrix provides a visual overview of the phase-dependent patterns learned by the model, where the color intensity represents the magnitude of the attention elements—darker colors indicate stronger correlations between qubits, while lighter colors indicate weaker correlations.

As illustrated in 
Fig.~\ref{fig:additional matrix}, for the 9-qubit system, the paramagnetic phase exhibits an attention pattern different from that shown in the main text. While the previous result featured the first qubit being weakly correlated with the others, here the seventh qubit instead displays weak correlations. For the 15-qubit system, multiple qubits (specifically the 5th, 8th, and 11th) exhibit similarly weak correlations with the rest of the system. These variations arise from differences in stochastic training processes, as all additional matrices were obtained from networks trained with identical hyperparameters but different initial parameters and training sets. This ensures that observed differences stem from stochastic initialization rather than changes in model architecture or training protocol.  

Despite minor variations in the specific qubits exhibiting weak correlations, the overall phase-dependent patterns remain consistent across independent training runs, demonstrating the robustness and generalizability of the model. The additional matrices thus provide further evidence that the hybrid quantum-classical attention model reliably captures distinctive correlation structures associated with different quantum phases, enabling stable and reproducible phase classification.

\bibliography{apssamp}% Produces the bibliography via BibTeX.

\end{document}